\documentclass[]{spie}  %>>> use for US letter paper
\usepackage{amsmath,amsfonts,amssymb}
\usepackage{graphicx}
\usepackage{bm}
\usepackage[hang]{subfigure}
\usepackage[colorlinks=true, allcolors=blue]{hyperref}

\title{SPRINT for WFAO systems}

\author[a,b]{Guido Agapito}
\author[a,b]{Cédric Plantet}
\author[c,d]{Cédric Taïssir Heritier}

\affil[a]{INAF -- Osservatorio Astrofisico di Arcetri, Largo E. Fermi 5, 50125, Firenze, Italy}
\affil[b]{ADaptive Optics National laboratory in Italy (ADONI)}
\affil[c]{DOTA, ONERA [F-13661 Salon cedex Air - France]}
\affil[d]{Aix Marseille University, CNRS, CNES, LAM, Marseille, France}

\authorinfo{Further author information: (Send correspondence to G.A.)\\G.A.: E-mail: guido.agapito@inaf.it}

\begin{document} 
\maketitle

\begin{abstract}
The calibration of future wide field adaptive optics (WFAO) systems requires knowledge of the geometry of the system, in particular the alignment parameters between the sub-apertures of the wavefront sensors (WFS), pupil and deformable mirror (DM) actuator grid. Without this knowledge, closed-loop operation is not possible and the registration must be identified with an error significantly smaller than the sub-aperture size to achieve the nominal performance of the adaptive optics system. Furthermore, poor accuracy in this estimation will not only affect performance, but could also prevent the closed loop from being stable.
Identification is not trivial because in a WFAO system several elements can move with respect to each other, more than in a SCAO system. For example, the pairing of the sub-aperture and the actuator grating on a DM conjugated to an altitude different from 0 can depend on the size of the pupil on the WFS, the exact conjugation of the DM, the position of the guide star and the field rotation. This is the same for each WFS/DM pair.
SPRINT, System Parameters Recurrent INvasive Tracking, is a strategy for monitoring and compensating for DM/WFS mis-registrations and has been developed in the context of single conjugate adaptive optics (SCAO) systems for the ESO Extremely Large Telescope (ELT).
In this work, we apply SPRINT in the context of WFAO systems with multiple WFSs and DMs, investigating the best approach for such systems, considering a simultaneous identification of all parameters or subsequent steps working on one DM at a time.

\end{abstract}


\section{INTRODUCTION}
\label{sec:intro}  % \label{} allows reference to this section

Many of the current and future astronomical instruments for 8, 30 and 40 m class telescopes are assisted by Adaptive Optics (AO) systems to improve image quality as close to the diffraction limit as possible\cite{2012ARA&A..50..305D}.
The most common AO system is the single conjugate, where the wavefront correction is driven by the measurement of the light coming from a  guide star (GS): this approach provides a very high correction at the position of the GS, which decreases rapidly in the field.
To overcome this limitation, a type of AO has been developed that provides image correction over an extended field of view: the so-called Wide Field Adaptive Optics (WFAO).
Several flavours of WFAO are developed for astronomical telescopes: Ground Layer Adaptive Optics (GLAO\cite{2016SPIE.9909E..36O,2018SPIE10703E..1GO,2022SPIE12185E..6MT}), Multi-Conjugate Adaptive Optics (MCAO\cite{2018ARA&A..56..277R,2014MNRAS.437.2361R,2021Msngr.182...13C,2021Msngr.185....7R}) and Multi-Object Adaptive Optics (MOAO\cite{2021Msngr.182...33H,2022SPIE12185E..4AC}).
These WFAO system comprise multiple wavefront sensors (WFS) and one or more Deformable Mirrors (DM).
In this context, as for SCAO systems\cite{2022SPIE12185E..5OV,2023aoel.confE..62H}, estimating the alignment parameters between the actuators and the WFSs is critical, especially in systems where mechanical flexures, rotations and other dynamic effects change this alignment during operation, and even when introduced intentionally\cite{CranneySPIE}.
Different approaches to the identification of these parameters were developed, for example the ones described in Refs. \citeonline{2012SPIE.8447E..2DK,2021MNRAS.504.4274H,Berdeu2024}.
One type of identification that differs from the previous ones is the illumination-based identification: this approach has been chosen at the LBT for FLAO/SOUL\cite{2010SPIE.7736E..09E,2023aoel.confE..80P} and at the VLT for ERIS\cite{2022SPIE12185E..08R}, to name but a few, and it is the only approach that works to measure pupil position in a system where the position of the pupil and of the actuators of the DM are not fixed (\emph{e.g.} ELT\cite{2023ConPh..64...47P} where M4\cite{2019Msngr.178....3V} is conjugated at about 600 m).
In this paper we have chosen the approach presented in Heritier et al.\cite{2021MNRAS.504.4274H} and we extend it to the WFAO case.

In the next section we present the mis-registration we expect in a WFAO system and we discuss about how they are seen by pairs of WFS and DM and the relationship with the global geometry.
Then, in Sec. \ref{sec:rec}, we show how the SPRINT approach is extended to the WFAO case and, in Sec. \ref{sec:sim}, we show a couple of examples from numerical simulations.
Finally, we present our conclusion in Sec. \ref{sec:conc}.

\section{Mis-registrations} 
\label{sec:misreg}

Apart from the higher order pupil distortion, the main parameters that SPRINT should estimate are shifts, rotation and magnification between the sub-apertures of the WFS and the actuators of the DM.
In a SCAO system the mis-registrations are between a single WFS and a single DM and their identification is direct, whereas in a WFAO system with multiple WFSs and DMs the mis-registration between pairs of WFS and DM gives a local and relative view of the overall geometry of the system which can be identified by an additional reconstruction step. 
This reconstruction is required to convert the local mis-registration geometry to the global one.
Local mis-registration geometry comprises:
\begin{itemize}
    \item X and Y shift between WFS sub-apertures and DM actuators.
    \item Rotation between WFS sub-apertures and DM actuators.
    \item Magnification between WFS sub-apertures and DM actuators (in this paper we consider one parameter for simplicity, but X and Y magnification or higher order distortion can also be considered).
\end{itemize}
Global mis-registration geometry comprises:
\begin{itemize}
    \item X and Y shift between WFS sub-apertures and pupil (note that, as mentioned in the Sec. \ref{sec:intro}, the alignment of the pupil to the sub-apertures of the WFS can only be retrieved by a flux-based method as will be done in MORFEO\cite{CapassoSPIE,LampitelliSPIE}).
    \item Rotation between WFS sub-apertures and pupil.
    \item Magnification WFS sub-apertures and pupil.
    \item X and Y shift between DM actuators and pupil (this type of mis-registration occurs in a system where the DM is not the pupil: e.g. a post-focal DM in an MCAO system).
    \item Rotation between DM actuators and pupil.
    \item Magnification DM actuators and pupil.
    \item X and Y shift between actual and nominal guide star position.
    \item Shift between actual and nominal guide star altitude (only for laser guide stars).
\end{itemize}
If one DM is the pupil (for example the ASM\cite{2010SPIE.7736E..2CR} of LBT or the DSM\cite{2014SPIE.9148E..45B} of VLT) the global mis-registration is simplified and some elements of the previous list merge together, for example \textit{X and Y shift between WFS sub-apertures and pupil} and \textit{X and Y shift between DM actuators and pupil} becomes \textit{X and Y shift between WFS sub-apertures and DM actuators} so the global geometry for this point is the same as the local one.
Instead, note that the global mis-registration geometry cannot be fully recovered by considering one DM at a time, because the identification of some elements of the global mis-registration geometry requires the knowledge of the full set of local mis-registrations, such as \textit{X and Y shift between actual and nominal guide star position} and \textit{X and Y shift between WFS sub-apertures and DM actuators}, which cannot be distinguished unless all DMs are considered.

\begin{figure}[h]
    \centering
    \includegraphics[width=0.8\columnwidth]{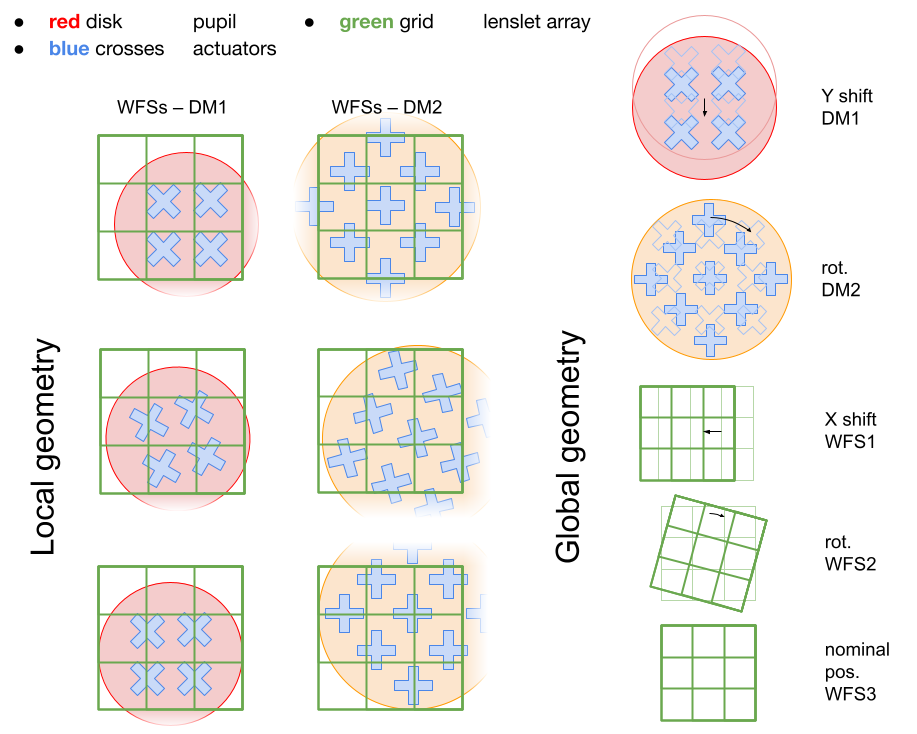}
    \caption{Diagram showing an example of mis-registration for a WFAO system with 3 WFSs and 2 DMs. One DM is also the pupil. For simplicity, no magnification is taken into account. Left and center, local mis-registration as seen by the WFS, right, global geometry.}
    \label{fig:misreg_geometry}
\end{figure}

An example of the relationship between local and global geometry is presented in Fig. \ref{fig:misreg_geometry}: we have 3 WFSs and 2 DMs, one in pupil and one conjugated at higher altitude.
The local geometry between each pair WFS -- DM is shown on the left and the central part of the figure, some shifts and rotations are visible, but it is not easy to determine the global geometry shown on the right.
The first DM is shifted downwards and this can be seen in the local geometry of all 3 WFSs with this DM, while the shift to the right of the first WFS is seen in the local geometry of this WFS with both DMs.
Rotations could have a more complex effect, since the centre of rotation depends on the shifts, as can be seen in the figure.
In particular, for the second DM, the pupil shifts are intrinsic due to the different line of sight of the WFSs.
Thus, rotation has local effects that are a combination of shift and rotation.
\section{Identification of global geometry}
\label{sec:rec}

The identification of the global geometry can be done by generalizing the SCAO-like identification presented in Ref. \citeonline{2021MNRAS.504.4274H}.
If we define as $D^{x,y}$ the Interaction Matrix (IM) between WFS $x$ and DM $y$ (a single mode per DM can be considered here and not the full IM), the global IM in the case of the example presented in Sec. \ref{sec:misreg} is:
\begin{equation}
    D = \left[
\begin{array}{cc}
D^{1,1} & 0\\
D^{2,1} & 0\\
D^{3,1} & 0\\
0 & D^{1,2}\\
0 & D^{2,2}\\
0 & D^{3,2}\\
\end{array}
\right]
\end{equation}
Note that the first (meta) row of the $D$ matrix corresponds to the estimation of the local geometry between first WFS and first DM\footnote{the second meta row corresponds to the second WFS, ... and the second meta column corresponds to the second DM.}, for this reason the second column is 0.
The same applies to other combinations.

As for the IM we can define a global sensitivity matrix.
If we define the sensitivity matrix between WFS $x$ and DM $y$ for the mis-registration $\alpha$ as:
\begin{equation}
    \Lambda^{x,y}(\alpha) = \left[ \frac{D^{x,y}(\alpha+\Delta \alpha)-D^{x,y}(\alpha-\Delta \alpha)}{2\Delta \alpha} \right] \; ,
\end{equation}
in a simplified case with 3 shifts, $\alpha_1$ and $\alpha_2$ on each one of the 2 DMs and $\alpha_3$ on one of the 3 WFSs, we have:
\begin{equation}
\Lambda = \left[
\begin{array}{ccc}
\Lambda^{1,1}(\alpha_1) & 0 & \Lambda^{1,1}(\alpha_3) \\
\Lambda^{2,1}(\alpha_1) & 0 & 0 \\
\Lambda^{3,1}(\alpha_1) & 0 & 0 \\
0 & \Lambda^{1,2}(\alpha_2) & \Lambda^{1,2}(\alpha_3) \\
0 & \Lambda^{2,2}(\alpha_2) & 0 \\
0 & \Lambda^{3,2}(\alpha_2) & 0 \\
\end{array}
\right] \; .
\end{equation}
Note that the zeros are located where the shifts have no effect, for example the first shift, $\alpha_1$ on the first DM is not seen on the local geometry between WFSs and DM no. 2, while third shift, $\alpha_3$, on the first WFS is not seen on WFS no. 2 and 3.

So following the same implementation as the one of a SCAO system the registration state $\bm{\alpha}$ can be computed iteratively as:
\begin{equation}
\hat{D}_k = \hat{D}(\bm{\hat{\alpha}}(k)) \; ,
\end{equation}
\begin{equation}
\Lambda_k = \Lambda(\bm{\hat{\alpha}}(k)) \; ,
\end{equation}
\begin{equation}
G_k = \mathrm{diag}(\hat{D}_k^{\dagger} \cdot D) \; , 
\end{equation}
\begin{equation}
\bm{\hat{\alpha}}(k+1) = \Lambda_k^{\dagger} \cdot (\hat{D}_k G_k^{-1} - D) \; ,  
\end{equation}
where $G$ is a gain matrix that model the sensitivity of the sensors, $k$ is the estimation step, $\hat{}$ is used for the estimated variables and $^{\dagger}$ is the pseudo-inverse.

\subsection{Interaction Matrices}
\label{sec:IMs}

\begin{figure}[h]
    \centering
    \subfigure[Geometry of the metapupil and the LGS footprint.\label{fig:MAVIS_DM3_geom}]
    {\includegraphics[width=0.4\columnwidth]{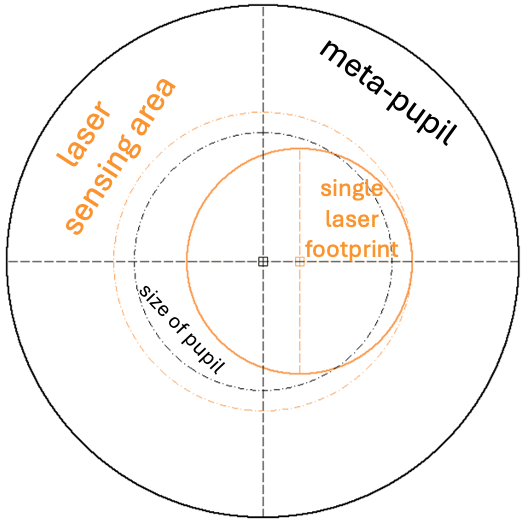}}
    \subfigure[Shape (wavefront) of a mode with high contrast where the sensing is performed.\label{fig:MAVIS_DM3_mode}]
    {\includegraphics[width=0.4\columnwidth]{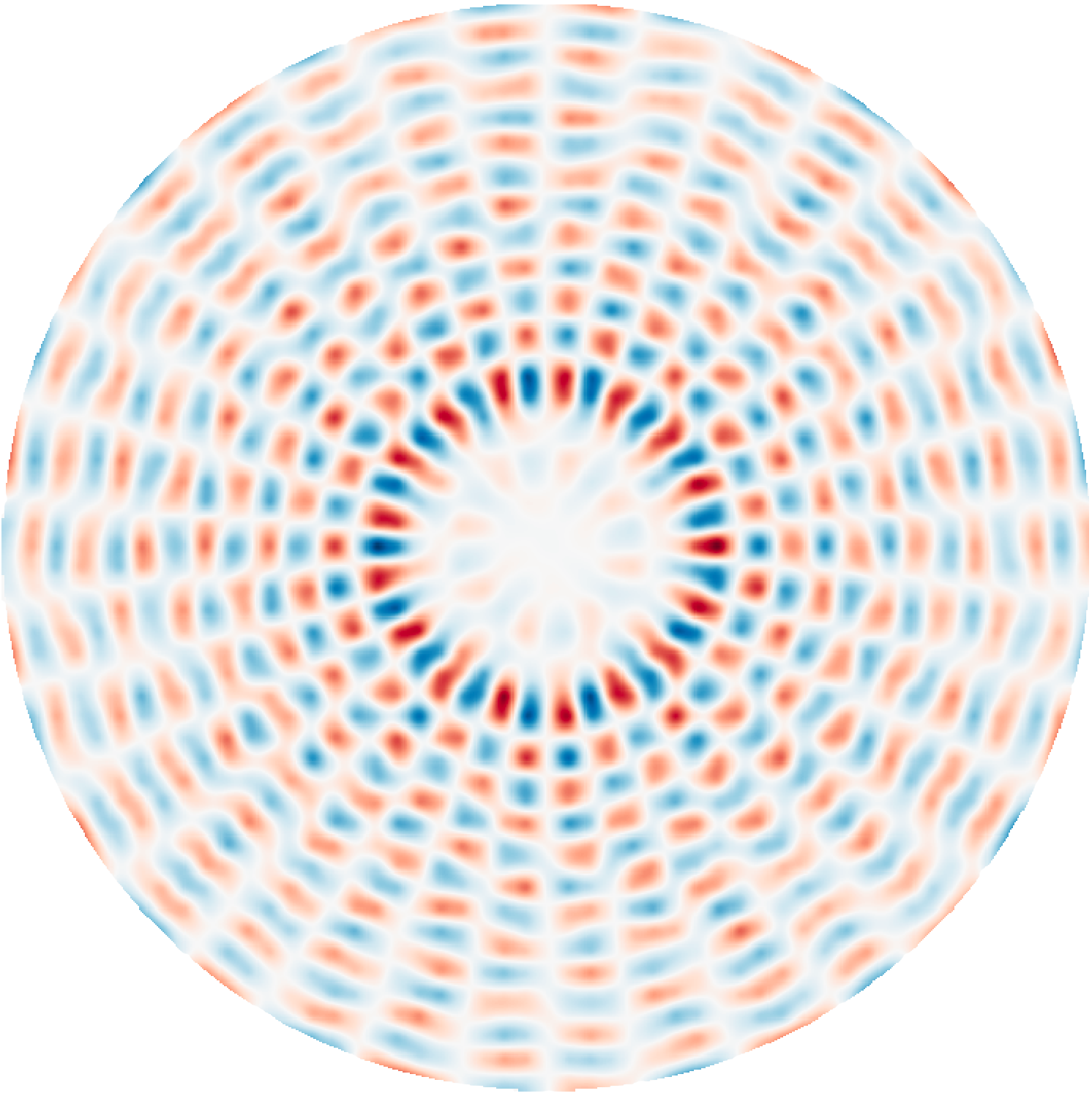}}
    \caption{Example of the third DM of MAVIS. Altitude is 13.5km, the LGS is pointing at 17.5 arcsec off-axis and the distance from the pupil of the LGS spot is 104 km (90 km with a zenith angle of 30deg).}
    \label{fig:MAVIS_DM3}
\end{figure}
The approach illustrated in the previous sections requires a set of measured IMs, one for each pair of WFS-DM. 
These IMs must be obtained on sky from the WFS measurements of a probe signal (sinusoidal or push-pull) on at least one mode per DM.
In this paper we do not focus on the aspect of acquisition on sky of the IM or the topic of modulation-demodulation of a probe signal, which has already been discussed in other papers, e.g. in Refs. \citeonline{2012SPIE.8447E..5MR,2012SPIE.8447E..2BP,2015aoel.confE..36E,2020A&A...636A..88E,2021MNRAS.504.4274H}.
Here we focus on the selection of the modes.
As it can be seen from the example presented in Sec. \ref{sec:misreg} the WFSs detect only part of the DMs conjugated in altitude: this means that the modes chosen for identification should take this into account.
Thus, if we consider the third DM of MAVIS\cite{2022SPIE12185E..20V,2022SPIE12185E..6PG} conjugated at 13.5 km, only a small part of the DM is detected by a single LGS as can be seen in Fig. \ref{fig:MAVIS_DM3_geom}.
Therefore, the shape of mode selected for on sky calibration of the IM has a high contrast where the sensing is performed as can be seen in Fig. \ref{fig:MAVIS_DM3_mode}. 

\section{Simulation}
\label{sec:sim}

\begin{table}[h]
\caption{Mis-registration status of the MAVIS example. Only the first WFS is present as the others are in the nominal registration.}
\label{tab:misreg}
\begin{center}
\begin{small}
	\begin{tabular}{|l|c|c|c|c|}
		\hline
		\textbf{mis-registration} & \textbf{WFS1} & \textbf{DM1} & \textbf{DM2} & \textbf{DM3}\\
		\hline
		\textbf{X shift [SA]} & -0.11 & 0 & 0.22 & 0\\
		\hline
		\textbf{Y shift [SA]} & 0.11 & 0 & 0.22 & 0\\
		\hline
		\textbf{rotation [deg]} & -0.05 & 0 & 0 & -0.1\\
		\hline
		\textbf{magnification [\%]} & 0 & 1 & 0 & 0\\
		\hline
	\end{tabular}
\end{small}
\end{center}
\end{table}
\begin{figure}[h]
    \centering
    \includegraphics[width=0.8\columnwidth]{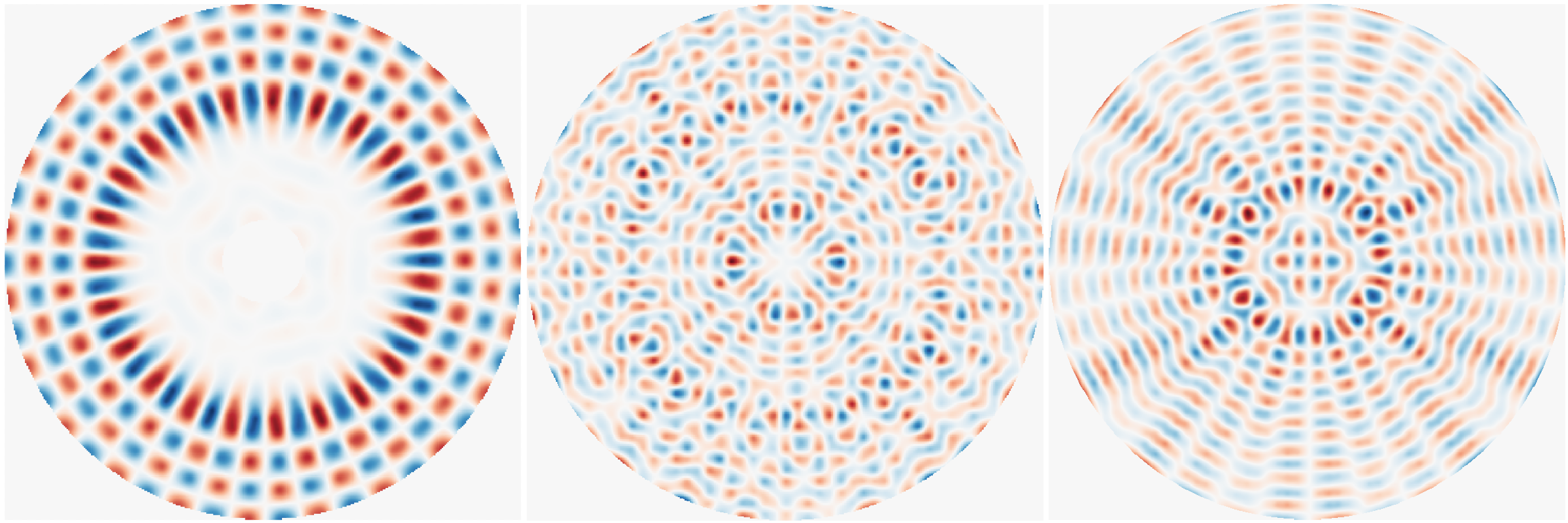}
    \caption{Shape (wavefront) of the mode selected for the calibration of the on-sky IM. Left, first DM (ground), center, second DM (conjugation altitude 6 km) and, right, third DM (conjugation altitude 13.5 km).}
    \label{fig:MAVIS_DMs_modes}
\end{figure}
\begin{figure}[h]
    \centering
    \includegraphics[width=0.5\columnwidth]{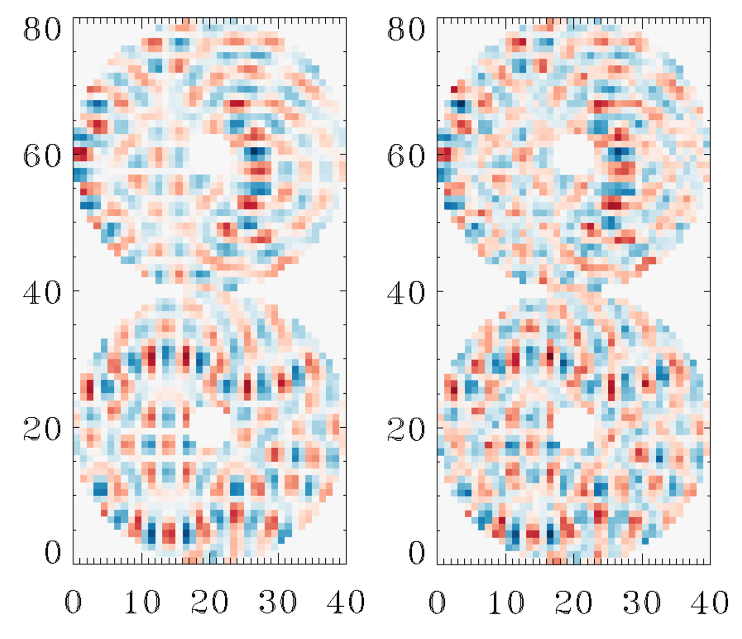}
    \caption{Example of X and Y signal from the IMs between the third DM and the first WFS of MAVIS: left, synthetic IM, right, on-sky measured (in simulation).}
    \label{fig:MAVIS_DM3_IM}
\end{figure}

In this section we present the result of a simulation done with PASSATA\cite{2016SPIE.9909E..7EA} of an AOF\cite{2018SPIE10703E..1GO} (GLAO) and MAVIS\cite{2022SPIE12185E..3LA} (MCAO) examples.

The mis-registration status is depicted in Tab. \ref{tab:misreg}.
In the case of AOF the mis-registration is the same except that there is only one DM.
An example of X and Y signals from the IMs used for the identification is shown in Fig. \ref{fig:MAVIS_DM3_IM}.

The identification is performed as described in the previous section, and we have used two sets of simulation data: a first set without atmospheric disturbances or noise, where the on-sky IMs are perfect, and a second set where a seeing of 0.87 arcsec on the line of sight is present and the WFSs are affected by noise.
In both cases the spots elongation due to the sodium profile is present.
The first set is used to evaluate the limits of this approach, while the second is used to obtain the precision when considering realistic conditions.

The modes selected for the DMs are shown in Fig. \ref{fig:MAVIS_DMs_modes}.
The probe signals used for the on-sky calibration of the IMs are a sinusoidal signal with an amplitude of 5 nm RMS, a frequency of 150, 153 and 156 Hz respectively on the 3 DMs and a duration of 10 seconds.
Note that in the case of AOF only the first DM is present and that we have limited the maximum number of iterations in the identification process to 5.

The AOF (GLAO) results can be summarized as:
\begin{itemize}
    \item No atmospheric disturbances or noise: the accuracy in the identification of the error is 0.001 SA, 0.003 deg and 0.005 diameter for shift, rotation and magnification respectively.
    \item Seeing 0.87 arcsec and noise: the accuracy in the identification of the error is 0.03 SA, 0.02 deg and 0.0004 diameter for shift, rotation and magnification respectively.
\end{itemize}

The MAVIS (MCAO) results can be summarized as:
\begin{itemize}
    \item No atmospheric disturbances or noise: the accuracy in identifying the error for the local geometry is 0.001 SA, 0.003 deg and 0.005 diameter for shift, rotation and magnification respectively.
    The error for the global geometry is slightly higher and in particular for the post-focal DMs there is a factor of about 2 for shift and rotation.
    \item Seeing 0.87 arcsec and noise: the accuracy in identifying the error for the local geometry is 0.03 SA, 0.03 deg and 0.005 diameter for shift, rotation and magnification respectively  (as for the AOF example) and, as seen in the previous point, there is a factor of 2 on the error in identifying the shift and rotations of the post-focal DMs.
\end{itemize}

Thus, although we have not studied the issue in depth, it appears that there is some cross-talk between shifts and rotations on post-focal DMs.
This could be related to the choice of modulated modes or to some specific aspect of the system geometry.
Nevertheless, the errors are small enough that no problems are expected with AO control and performance in a MCAO system, as can be seen in Refs. \citeonline{2020SPIE11448E..2SA,2022SPIE12185E..3LA}.

\section{CONCLUSION}
\label{sec:conc}

In this paper we have shown how SPRINT can be used effectively to estimate mis-registrations in WFAO systems with small estimation estimation errors of shift, rotation and magnification that are $<$0.1SA, $<$0.1deg, $<$1\% respectively.

Interestingly, the amplitude of the modulated modes is small enough (5nm amplitude) to have a negligible effect on performance, and the estimation error is not worse when the AO correction is poor, as in a GLAO system: this is the advantage of a linear WFS such as the Shack-Hartmann sensor (in contrast the pyramid WFS requires a modulation signal with a larger amplitude when the AO correction is poor\cite{2015aoel.confE..36E,2020A&A...636A..88E}).

The results are still excellent for the MCAO example we presented, although we have seen that the identification of mis-registration on the post-focal DMs is more difficult.
Future work will be to study this case in more detail, exploring different combinations of modulated modes and/or different reconstruction approaches.

%\acknowledgments % equivalent to \section*{ACKNOWLEDGMENTS}       
%XXX.  

% References
\bibliography{report} % bibliography data in report.bib
\bibliographystyle{spiebib} % makes bibtex use spiebib.bst

\end{document}